\documentclass[epsfig,12pt]{article}

\usepackage{latexsym}
\usepackage{relsize}
\usepackage{geometry}
 \usepackage{amsmath, amssymb, amscd, xypic, graphicx}
\usepackage{slashed,color}
\usepackage{epstopdf}
\input xy
\xyoption{all}

 \usepackage{fancyhdr}                   
\def\beq{\begin{equation}}
\def\eeq{\end{equation}}
\def\beqn{\begin{eqnarray}}
\def\eeqn{\end{eqnarray}}

\newcommand{\ntwo}{${\mathcal N}=2\,$}

\newcommand{\nzt}{${\mathcal N}=(0,2)\,$}
\newcommand{\cpn}{CP$(N-1)\,$}

\newcommand{\sma}{\left(\begin{smallmatrix}}
\newcommand{\smaa}{\end{smallmatrix}\right)}

\newcommand{\gsim}{\lower.7ex\hbox{$
\;\stackrel{\textstyle>}{\sim}\;$}}
\newcommand{\lsim}{\lower.7ex\hbox{$
\;\stackrel{\textstyle<}{\sim}\;$}}

\begin{document}


\begin{titlepage}

\begin{flushright}
FTPI-MINN-10/24, UMN-TH-2916/10\\
\end{flushright}

\vspace{1cm}

\begin{center}
{  \Large \bf  Perturbative Aspects of Heterotically Deformed\\[2mm]
 CP\boldmath{$(N-1)$} Sigma Model. I }
\end{center}

\vspace{0.8cm}

\begin{center}
{\large
Xiaoyi Cui$^{\,a}$ and M.~Shifman$^{\,b}$}
\end {center}

\vspace{1mm}

\begin{center}

$^{a}${\it   Physics Department, University of Minnesota,
Minneapolis, MN 55455, USA}\\[2mm]
$^b${\it  William I. Fine Theoretical Physics Institute,
University of Minnesota,
Minneapolis, MN 55455, USA\,\footnote{Permanent address.}\\[-1mm]
{\small and} \\
 Jefferson Physical Laboratory, Harvard University, Cambridge, MA 02138, USA }

\end{center}

\vspace{1cm}

\begin{center}
{\large\bf Abstract}
\end{center}

\vspace{1.5cm}
In this paper we begin the study of renormalizations in the heterotically deformed $\mathcal{N}=(0,2)$ CP$(N-1)$ sigma models. In addition to the coupling constant $g^2$ of the undeformed $\mathcal{N}=(2,2)$ model, there is the second coupling constant $\gamma$ describing the strength of 
the heterotic deformation. We calculate both $\beta$ functions, $\beta_g$ and $\beta_\gamma$ 
at one loop  determining the flow of $g^2$ and $\gamma$. Under a certain choice of the initial conditions, the theory is asymptotically free.
The $\beta$ function for the ratio
$\rho =\gamma^2/g^2$ exhibits an infrared fixed point at $\rho=1/2$.
Formally this fixed point lies outside the validity of the one-loop
approximation. We argue, however, that the fixed 
point at $\rho =1/2$ may survive to all orders.
The reason is the enhancement of symmetry -- emergence of a chiral fermion flavor
symmetry in the heterotically deformed Lagrangian -- at $\rho =1/2$.
Next we argue that  $\beta_\rho$
 formally obtained at one loop, is  exact to all orders in the
 large-$N$ (planar) approximation. Thus, the fixed point at $\rho =1/2$
 is definitely the feature of the model in the large-$N$ limit.

\end{titlepage}

\newpage

\tableofcontents

\newpage

\section{Introduction}

Two-dimensional  CP$(N-1)$ models  
emerged as effective low-energy theories on
the worldsheet of non-Abelian strings in a class of 
four-dimensional \ntwo\, gauge theories~\cite{HT1,ABEKY,SYmon,HT2} (for reviews see  \cite{Trev}). 
Deforming these models in various ways (i.e. breaking supersymmetry down to ${\mathcal N}=1$)
one arrives at heterotically deformed \cpn models \cite{EdTo,SY1,BSY1,BSY2}
(heterotic \cpn models for short),
a very interesting and largely unexplored class of models characterized by two coupling constants:
the original asymptotically free coupling and an extra one describing the strength of the heterotic deformation.
These two-dimensional models are of importance on their own,
since they exhibit highly nontrivial dynamics, with a number of phase transitions. This fact was 
recently revealed \cite{largen} in the
large-$N$ solution of the model (see also \cite{KMV}).

Our current task is to analyze perturbative aspects of the heterotic $\mbox{\cpn}$ models.
General aspects of perturbation theory in the $\mathcal{N}=(0,2)$ models were discussed 
by Witten \cite{Witten05}. We will study particular renormalization properties and calculate 
the $\beta$ functions in the CP$(N-1)$ models heterotically deformed in a special way.
In this first paper of a series we will focus on one-loop effects 
and demonstrate that both couplings of the model
enjoy asymptotic freedom (AF). Moreover, we observe a special fixed-point regime in the 
infrared (IR) domain and argue that it holds beyond one loop.

Written in components, the  Lagrangian of the heterotic CP(1) model takes the form \cite{SY1}
 \beq
 {\mathcal{L}}_{(2,2)}
=
 G\left\{
\partial^\mu\phi\partial_\mu\phi^\dagger+i\bar{\psi}\slashed{\partial}\psi -
{2i}  \,\frac{1}{\chi}\, \bar{\psi}\gamma^\mu\psi\,\phi^\dagger\partial_\mu\phi-\frac{2}{\chi^2}\,\psi_L^\dagger\psi_L\,\psi_R^\dagger\psi_R 
\right\}\,,
\label{N22model}
 \eeq
 and\,\footnote{The sign in front of the term 
 $\zeta_R^\dagger\, \zeta_R \psi_L^\dagger\psi_L$ in (\ref{N02model})
is opposite to that in \cite{SY1} due to a typo in \cite{SY1}. Also notice that the definition of $\gamma$ in this paper corresponds to $\gamma g^2$ in \cite{SY1}. The reason for 
this rescaling of the deformation parameter  compared to \cite{SY1} is that $g^2$ and $\gamma^2$ 
as defined here are the genuine loop expansion parameters.}
\beqn
&&
{\mathcal L}_{(0,2)}= 
\zeta_R^\dagger \, i\partial_L \, \zeta_R  + 
\left[\frac{\gamma}{g^2} \, \zeta_R  \,R\,  \big( i\,\partial_{L}\phi^{\dagger} \big)\psi_R
+{\rm H.c.}\right] 
\nonumber
\\[3mm]
&&
+\frac{|\gamma |^2}{g^2} \left(\zeta_R^\dagger\, \zeta_R
\right)\left(R\,  \psi_L^\dagger\psi_L\right)
+
G\, \left\{
\frac{2 |\gamma |^2}{g^2\chi^2}\,\psi_L^\dagger\,\psi_L \,\psi_R^\dagger\,\psi_R
\right\}\,,
\label{N02model}
\eeqn
where $G$ is
the K\"ahler  metric on the target space, 
\beq
G =
\frac{2}{g^2\,\chi^{2}}\,,
\label{kalme}
\eeq
$R$ is the Ricci tensor,
\beq
 R =\frac{2}{\chi^2}\,,
\label{Atwo}
\eeq
and we use the notation 
\beq
\chi \equiv 1+\phi\,\phi^\dagger\,.
\label{chidef}
\eeq
The coupling $g^2$ enters through the metric, while the deformation coupling $\gamma$
appears in Eq.~(\ref{N02model}). Both couplings can be chosen to be 
real.\footnote{However, in Secs.~\ref{section2} and~\ref{section3} 
 we will treat $\gamma$ as a complex coupling in analyzing generalized U(1) 
 symmetries and for similar purposes.}
Setting $g^2$ to be real means we will not consider the topological term 
$\varepsilon_{\mu\nu}\partial^\mu\phi\partial^\nu\phi^\dagger$ 
in this paper. To make $\gamma$  real, one should perform a phase rotation of the bosonic field $\phi$ absorbing the phase of $\gamma$ (see the first line in Eq.~(\ref{N02model})).
In fact, this corresponds to a kind of $R$-symmetry, as the reader will see from the $\mathcal{N}=(0,2)$ superfield formalism In Sec.~\ref{section2}.
Our main results are presented by the following expressions for the
one-loop $\beta$ functions:
\beqn
\beta_g &\equiv& \frac \partial{\partial\text{ln}\mu}g^2(\mu)
=
-\frac{g^4}{2\pi} + ...
\label{betafsp}
\\[2mm]
\beta_\gamma &\equiv& \frac \partial{\partial\text{ln}\mu}\gamma(\mu)
=
\frac{\gamma}{2\pi}\left(\gamma^2-g^2\right) + ...
\label{betafs}
\eeqn
where ellipses stand for two-loop and higher-order terms. The heterotic deformation does not
affect $\beta_g$ which stays the same as in the ${\mathcal N}=(2,2)$ CP(1) model.
Among other results, we calculate the law of running of the ratio $$\rho=\frac {\gamma^2}{g^2}\,$$
see Eq.~(\ref{47}). If at any renormalization point, in particular,
in the ultraviolet (UV) limit, $\rho$ is chosen to be smaller than $1/2$, in the IR limit 
it runs to $\rho\to 1/2$, which is the fixed point for this 
parameter. With $\rho\leq 1/2$, the theory is asymptotically free. Analogs of Eqs.~(\ref{betafsp}) 
and (\ref{betafs})  in the heterotic CP$(N-1)$ model with arbitrary $N$
are presented in (\ref{53}) and  (\ref{55}).

The paper is organized as follows.
In Sec.~\ref{section2}, we describe the model using both superfield language and component field language. We show that the $\mathcal{N}=(0,2)$ deformation structure is quite unique in certain sense. In Sec.~\ref{section3}, we show that by symmetry analysis, the renormalization structure of the deformation term is  constrained. Conceptually it gives partially the answer for why the deformation term does not get renormalized at one-loop level. In Sec.~\ref{section4} we describe the linear background field method and calculate the one-loop $\beta$ function for $g^2$ as an example. This will be useful in the two-loop calculations shown in our subsequent paper. In Sec.~\ref{section5} we calculate the Z-factor wave-function renormalizations for the field $\zeta_R$ and $\psi_R$. In Sec.~\ref{section6} we calculate the one-loop $\beta$ function for $\gamma$, and thus verify our expectation in Sec.~\ref{section3}. In Sec.~\ref{section7} we discuss the running of the two couplings. We find the region that is good for perturbative calculation, and we find an IR fixed point of $\rho$ that exists universally at one-loop order. In Sec.~\ref{section8} we generalize the result to CP$(N-1)$ model. We show that a factorization of $\beta_\rho$ survives in all loop orders. In the subsequent paper we will discuss the result from the two-loop calculation.

\section{\boldmath{$\mathcal{N}=(0,2)$} CP(1) sigma model}
\label{section2}

Here we will briefly review
basics of the heterotic CP(1) model. The metric (\ref{kalme}) is the Fubini--Study metric
of the two-dimensional sphere $S^2$.
It is not difficult  to see that the Lagrangian $ {\mathcal{L}}_{(2,2)} +  {\mathcal{L}}_{(0,2)}$ is invariant with respect to the joint U(1) rotations of 
the fields $\phi$ and $\psi$, and, in addition,  invariant under the following 
nonlinear transformations:
\beqn
&&
\phi\rightarrow\phi+\alpha+\alpha^\dagger\phi^{2}\,,\qquad \phi^\dagger\rightarrow\phi^\dagger+\alpha^\dagger+\alpha\phi^{\dagger2},\nonumber\\[3mm]
&&\psi\rightarrow \psi+2\alpha^\dagger\phi \psi\,,\qquad
\bar{\psi}\rightarrow \bar{\psi}+2\alpha\phi^\dagger\bar{\psi}.
\label{8}
\eeqn
Here $\alpha$ is a complex constant.
The above transformations tell us that $\phi$ transforms as the coordinate of
the target manifold (a 2-sphere), and the fields $\psi$ and $\partial^\mu\phi$ 
transform as the tangent vectors.

Supersymmetry of this model is best understood via its superfield description. 
The $\mathcal{N}=(0,2)$ model description can be obtained by either integrating out the Grassmannian variables $\theta_L$ and $\theta_L^\dagger$ \cite{SY1}, or by constructing the model from $\mathcal{N}=(0,2)$ superfields.\footnote{Our conventions, as well as the $\mathcal{N}=(0,2)$ superspace, are described in Appendix A.} We will follow the second way, for reasons which will become  clear shortly.
We define the left and right derivatives as follows.
\beq
\partial_L=\partial_t+\partial_z\,,\qquad \partial_R=\partial_t-\partial_z\,.
\eeq 
The shifted space-time coordinates that satisfy the chiral condition are defined as
\beq
x^0_L=t+i\theta_R^\dagger\theta_R\,,\qquad x^1_L=z+i\theta_R^\dagger\theta_R\,.
\eeq 
We start from  two chiral $\mathcal{N}=(0,2)$ superfields $A$ and $B$,
\beqn
&&
A(x^\mu_L,\theta_R)=\phi(x^\mu_L)+\sqrt{2}\theta_R\psi_L(x^\mu_L)\,,\nonumber\\[3mm]
&&
B(x^\mu_L,\theta_R)=\psi_R(x^\mu_L)+\sqrt{2}\theta_R F(x^\mu_L)\,.
\label{9}
\eeqn
The supersymmetry transformations are
\beqn
&&\delta_R\phi=\sqrt{2}\epsilon_R \psi_L\,,\qquad \delta_R
\psi_L=-\sqrt{2}i\epsilon_R^\dagger \partial_L\phi\,,\nonumber\\[3mm]
&&\delta_R \psi_R=\sqrt{2}\epsilon_R F\,,\qquad \delta_R F=-\sqrt{2}i\epsilon_R^\dagger\partial_L\psi_R\,.
\label{n02transformation}
\eeqn
In terms of these \nzt superfields, it is not difficult to show that the  Lagrangian 
\beq
\mathcal{L}_{(2,2)}=\frac 2{g^2}\, \int d^2\theta_R \left\{\frac{i A^\dagger \partial_R A-iA\partial_R A^\dagger}{1+A^\dagger A}+\frac{2B^\dagger B}{(1+ A^\dagger A)^2}\right\} 
\rule{0mm}{9mm}
\label{11}
\eeq
identically reproduces Eq.~(\ref{N22model}).

Needless to say,  this Lagrangian is $\mathcal{N}=(0,2)$ invariant, by construction. 
Next, we must 
show that it is target-space invariant. 
The global U(1) symmetry of Eq.~(\ref{11}) is obvious. As far as the nonlinear transformations Eq.~(\ref{8})
are concerned,
 for the first term we have 
 \beq
 \delta_\alpha \, \frac{A\partial_R A^\dagger+A^\dagger \partial_R A}{1+ A^\dagger A}\propto
 \left(  \alpha \partial_R A^\dagger+\alpha^\dagger \partial_R A\right) ,
\rule{0mm}{9mm}
 \label{12}
 \eeq 
 which is a combination of holomorphic and antiholomorphic functions, and thus should vanish in
 the action.
Moreover, one can show that the second term is invariant by itself, i.e., 
\beq
 \delta_\alpha \, \frac{B^\dagger B}{(1+ A^\dagger A)^2} = 0\,.
\eeq
The relative coefficient between the first and the second terms
in Eq.~(\ref{11}) is unity because of the $\mathcal{N}=(2,2)$ symmetry 
which is implicit in Eq.~(\ref{11}). 
One can, of course, introduce another 
 coupling constant in front of the term quadratic in the $B$ field, but 
 then one can always absorb such constant in the normalization of $B$. 
 
Thus, with the given set of fields,
starting from \nzt supersymmetry,
 we get an enhanced $\mathcal{N}=(2,2)$ supersymmetry ``for free."  This phenomenon is
  similar to the  $\mathcal{N}=(1,1)$ case, as shown in \cite{Zumino79} (see also \cite{wittenold}).

This gives us a hint on the way of 
converting the $\mathcal{N}=(2,2)$ model into its heterotic $\mathcal{N}=(0,2)$ extension. Namely, one should add an interaction of the field $B$ with $A$ and $A^\dagger$ without involving
the field $B^\dagger$. Taking into consideration the target space symmetry, one can come up with the following deformation Lagrangian:
\beq
\Delta \mathcal{L} =-\int d^2\theta_R\, \frac{4\gamma}{g^2}\,  \frac{ B A^\dagger}{1+ A^\dagger A}+ {\rm H.c.}\,.
\label{14}
\eeq
The coupling $\gamma$ has dimension $m^{1/2}$, and must be viewed as
a complex Grassmann number. 
Now, in components, the deformation Lagrangian takes the form  
\beq
\Delta  \mathcal{L} =\left[ \gamma\,  G(i\partial_L\phi^\dagger)\psi_R+
{\rm H.c.}\right]+|\gamma|^2(G \psi_R^\dagger\psi_R)\,.
\label{langrangian2}
\eeq
The above construction, however, suffers from the fermion number nonconservation. The very 
last step which fixes this problem is  promoting $\gamma$ to a dynamic superfield. To this end we introduce a chiral 
superfield\,\footnote{Warning: the definition of the superfield $\mathcal{B}$ in this paper is slightly different
from that in \cite{SY1}.}
\beq
\mathcal{B}=\zeta_R+\sqrt{2}\theta_R\mathcal{F}\,,
\label{15}
\eeq
and then replace $\gamma$ by
\beq
 \gamma \to \gamma \, \mathcal{B}\,.
 \label{15p}
\eeq
As a result, the deformation Lagrangian takes the form
\beqn
\Delta \mathcal{L}
&=&
\int d^2\theta_R \left\{\left[ -\frac{4\gamma}{g^2}\,\frac{ \mathcal{B} B A^\dagger}{1+A^\dagger A}+ {\rm H.c.}
\right]+ 2 \mathcal{B}^\dagger\mathcal{B}\right\}
\nonumber\\[3mm]
&=& i\zeta_R^\dagger\partial_L\zeta_R+
\left[\gamma \, \zeta_R \, G(i\partial_L\phi^\dagger)\psi_R+
{\rm H.c.}
\right]  \nonumber\\[3mm]
&&+|\gamma|^2(\zeta_R^\dagger\zeta_R)(G \psi_R^\dagger\psi_R)
+|\gamma|^2G^2\psi_L^\dagger\psi_L\psi_R^\dagger\psi_R\,.
\label{17}
\eeqn
The target space invariance can be verified at the level of superfields, 
\beq
 \delta_\alpha \, \frac{\mathcal{B}B A^\dagger}{1+ A^\dagger A} \propto \alpha^\dagger  \mathcal{B}B\,,
 \label{deltagamma}
\eeq
which vanishes in the action. Needless to say, the $S^2$ target space transformations of 
${\mathcal B}$ are trivial, $\delta {\mathcal B}=0$.

Finally, we comment on the absorption of the U$(1)$
phase of the coupling $\gamma$ in this context. This means 
that whenever $\gamma=|\gamma| e^{i\omega}$, we can always 
redefine $\gamma\to \gamma e^{-i\omega}$ and $A^\dagger \to A^\dagger e^{i\omega}$ supplemented by a $\theta_R$ rotation. So $\gamma$ can always 
be kept real. Notably, this rotation only involves bosons. This is because, 
we can assign a proper U$(1)$ charge to $\theta_R$ to keep $\psi_L$ 
U$(1)$-neutral. Thus the rotation actually corresponds to a kind of $R$-symmetry. It is anomaly-free.

\section{Global symmetries and renormalization structure }
\label{section3}

In this section we analyze the renormalization structure of the heterotic CP(1) model. 
In the deformed model we have two dimensionless couplings, $g$ and $\gamma$. 
The first one is related to  geometry of the target space, and the second one
parametrizes the strength
of  the heterotic deformation. The first question to ask at one-loop level is that
whether or not  there is mixing between these two couplings. In addition, we
should verify that other structures, which are absent in Eq.~(\ref{11}) and Eq.~(\ref{17}),
do not show up as a result of loop corrections.
What can be said from  analyzing  the symmetries of the model at hand? We will
follow the line of reasoning  similar to the proof 
\cite{Seiberg93} of nonrenormalization theorems for superpotentials in four dimensions.
We start from the U(1) symmetries.\footnote{The coupling constant $\gamma$ 
can and will be ascribed U(1) charges, as in \cite{Seiberg93}. }
There are three generalized U(1) symmetries listed in Table~\ref{u1sym},
with the corresponding charge assignments. 
\begin{table}
\begin{center}
\begin{tabular}{| c || c | c | c |}
\hline
Fields & U$(1)_1$ & U$(1)_2$ & U$(1)_3$ \\[2mm] \hline
$\gamma$ & $1$ & $0$ & $-\frac 12$ \\[2mm]
$A$ & $0$ & $1$ & $-\frac 12$ \\[2mm]
$B$ & $0$ & $1$ & $\frac 12$ \\[2mm]
$\mathcal{B}$ & $-1$ & $0$ & $-\frac 12$ \\[1mm]\hline
\end{tabular}
\end{center}
\caption{\small Generalized U(1) symmetries of the heterotically deformed CP(1) model.}
\label{u1sym}
\end{table} 
Using these charge assignments one can show that the only nontrivial combination 
of $\gamma$, $ A^\dagger$, $B$, and $\mathcal{B}$ invariant 
under all U(1)  symmetries is
 given by $\gamma\mathcal{B}B A^\dagger$. 
 
 Since $\Delta {\mathcal L}$ presents the integral over $d\theta_R$ and $d\theta_R^\dagger$,
 (unfortunately), it is not analogous to  $F$ terms in conventional supersymmetric
  Lagrangians.  Hence, with radiative corrections included,
$\Delta {\mathcal L}$ can possibly contain U(1)-neutral pairs 
such as $ A^\dagger A$, $B^\dagger B$, $\mathcal{B}^\dagger\mathcal{B}$ 
and the modulus of the coupling
 $|\gamma|^2$. Due to dimensional reasons,   $B^\dagger B$ could only 
 show up in the 
 combination of the type ${ B^\dagger B}/{ \mathcal{B}^\dagger \mathcal{B} }$, which, obviously, cannot happen in loops
 provided our theory is properly regularized in the infrared domain (and, of course, 
it will be regularized both in the IR and UV
domains). Its absence can also be verified by analyzing a fermionic SU$(2)$ 
symmetry discussed in Sec.~\ref{section5}.
 
As for $ A^\dagger A$, this term  is constrained by the 
 non-linear target space symmetry. 
 The only allowed form is the 
 combination $$\frac {B A^\dagger}{1+ A^\dagger A}$$ and its Hermitian conjugate. This shows that the 
 structure of the deformation is quite unique. 
  Corrections in $g^2$ are strongly constrained too, as we will see momentarily.
 Thus, the corrections that can show up in loops are 
 $O(\gamma^2)$,  $O(\gamma^2 g^2)$,  $O(\gamma^4)$, and so on.

 The latter circumstance becomes clear if we take into consideration that
 the operator $R \partial_L\phi^\dagger \psi_R$ is, in fact, the superconformal anomaly of the undeformed theory
 \cite{SY1}, 
 \beq
 J_{sc,L}=\frac{-i\sqrt{2}}{2\pi}\frac{\partial_L\phi^\dagger \psi_R}{\chi^2}\,.
 \label{scaut}
 \eeq 
 Since the supercurrent is conserved, $R \partial_L\phi^\dagger \psi_R$ receives no corrections in 
 $g^2, \,\, g^4,$ etc. Certainly, on general grounds we can not exclude
 corrections of the type $|\gamma |^2 g^2$, but these can show up only in the second and higher loops.
 Therefore, at one loop the corrections to $\gamma$
 come from $O(|\gamma|^2)$ one-particle irreducible correction to the vertex, and the $Z$ factors of the fields of which the deformation term is built.
 Calculation of the $Z$ factors is carried out in Sect.~\ref{section5}.
 We check all  assertions made above on general grounds by explicit calculation of relevant diagrams
  in Sect.~\ref{section6}.

\section{Background field method. One-loop $\beta$ function for $g^2$}
\label{section4}

We will see shortly that $\beta_g$ is unaffected by the heterotic deformation at one loop. Before passing to direct calculations of the $Z$-factors for the 
fields $\zeta_R$, $\psi_R$, let us outline 
generalities of  
the background field method \cite{rev1}.\footnote{From Eq.~(\ref{11}) 
it is obvious that the $Z$ factor for the superfield $A$ is unity.}
The full power and efficiency of this method will become clear
in two-loop calculations~\cite{CS2}. Here will just ``calibrate" it in the simple problem
of the one-loop $\beta$ function $\beta_g$. The result is well-known, of course 
\cite{Polyakov:1975rr}; therefore, this exercise \cite{rev1}
is  needed mostly to introduce necessary conventions and notations.

Each field is decomposed into a background part (large part) and a quantum part (small 
fluctuations to be integrated over).
One can simplify calculations by making a particularly convenient choice of the background
fields, which may vary from problem to problem.  Then one expands the Lagrangian 
in the given background, in terms of quantum fields. Linear terms can be omitted.
Quadratic terms determine one-loop corrections.
Cubic and higher order terms are needed for two-loop calculations, i.e. for higher orders.
The two-loop calculations will be presented in a forthcoming paper \cite{CS2}. 
Note that, generally speaking, the background 
field does not have to respect the symmetry of the theory. However, the full symmetry
 must certainly be present in the final answer 
for ${\mathcal L}_{\rm eff}$.

After this remark we proceed to an  instructive one-loop calculation
of $\beta_g$. 
The split of the fields is
\beq
\phi \to \phi_0 + q
\eeq
where $q$ represents the quantum part to be integrated over in loops. For the background part we choose
in the problem at hand
\beq
\phi_0 = f e^{-ikx}\,,\qquad  \phi_0^\dagger = f^\dagger e^{ikx}\,,\qquad \phi_0^\dagger \phi_0 = f^\dagger f\,,
\label{fri4}
\eeq
where $f$ is a constant, and at the very end (after separating all terms quadratic in $k$) we can tend $k\to 0$.
The fermion field is quantum, by definition. The background field expression for the Lagrangian is
\beq
\mathcal{L}_0 = \frac{2}{g_0^2} \,\frac{k^2\, f^\dagger f}{(1+ f^\dagger f )^2}\,,
\label{fri5}
\eeq
where $g_0$ is the bare coupling (i.e. the coupling constant at the UV cut-off
$M_{\rm uv}$.
We will  concentrate on the boson contribution because the fermion loop at this level is 
finite and, thus, does not contribute to  the $\beta$ function, see Appendix B. The results in this appendix
will be used in two-loop calculations.

Expanding the Lagrangian up to the second order in the quantum fields, we obtain
\begin{eqnarray}
\mathcal{L}_{\phi(1)}&=&(G^{1,0}_0 q+G^{0,1}_0 q^\dagger )\partial_{\mu}\phi_0\partial^{\mu} \phi^\dagger_0+G_0(\partial_{\mu}q\partial^{\mu} \phi^\dagger_0+\partial_{\mu}\phi_0\partial^{\mu} q^\dagger)\,,\nonumber\\[3mm]
\mathcal{L}_{\phi(2)}&=&(G^{2,0}_0q^2+G^{0,2}_0q^{\dagger 2}+G^{1,1}_0q q^\dagger)\partial_{\mu}\phi_0\partial^{\mu}\phi^\dagger_0
\nonumber\\[3mm]
&+&
(G^{1,0}_0q+G^{0,1}_0 q^\dagger)(\partial_{\mu}q\partial^{\mu}\phi^\dagger_0+\partial_{\mu}\phi_0\partial^{\mu} q^\dagger)
+G_0\partial_{\mu}q\partial^{\mu} q^\dagger
\label{fri6}
\end{eqnarray}
where 
\beq
\left.
G^{i,j}_0\equiv\frac{1}{i!j!}\, \frac{\partial^i}{\partial\phi^i}\frac{\partial^j}{\partial\phi^{j \dagger }}
\frac{2}{g^2_0\,(1+\phi\phi^\dagger)^2}
\right|_{\phi_0,\phi^\dagger_0}\,,
\label{defineG}
\eeq
 and the metric $G_0$ coincides with the quantity $G_0^{0,0}$ defined in Eq.~(\ref{defineG}). In what follows we may drop the subscript $0$ on some quantities, whose meaning is obvious from the context.
In the first line in Eq.~(\ref{fri6}) extra source terms (irrelevant for what follows) may be necessary
 in order to discard $\mathcal{L}_{\phi(1)}$ because of its proportionality to the equation of motion for the given background field.\footnote{In two-loop calculations one should also add to (\ref{fri6}) the fermion term ${\mathcal L}_\psi$ which can be read off from 
 Eq.~(\ref{N22model}).}
 
 In calculating the graphs which determine the $\beta$ function one should keep in mind that
 they are divergent both in the ultraviolet and infrared.
 To regularize the UV divergence we will use dimensional reduction working in $D=2-\epsilon$ dimensions.
To regularize the IR divergence  we introduce a small mass term for the quantum field (see \cite{rev1}),
\beq
{\mathcal L}_m = G_0 \left(- m_0^2 q^\dagger q - m_0\bar\psi\psi\
\right)\,.
\label{mon8}
\eeq
 At the very end both regularizing parameters, $m_0$ and $\epsilon$ are supposed to tend to zero.
 In logarithmically divergent graphs the following correspondence takes place at one loop:
 \beq
 \frac{1}{\epsilon}\, \leftrightarrow \,\text{ln}\frac{M_{\rm uv}}{m_0}\,,
 \label{mon9}
 \eeq
where the left-hand side represents dimensional regularization, while the right-hand side the Pauli--Villars regularization; $M_{\rm uv}$ is the mass of the Pauli--Villars regulator.

\begin{figure}
\begin{center}
\includegraphics[width=3in]{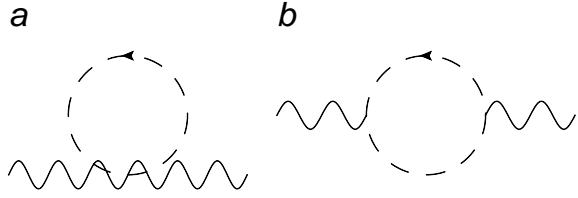}
\end{center}
\caption{\small Boson loops contributing to the one-loop $\beta$ function. The wavy lines stand for the background field
$\partial_\mu\phi_0$ while the dashed lines for $q$. }
\label{fig1}
\end{figure}

Now we can pass to one-loop calculation. 
The relevant boson diagrams are depicted in Fig.~\ref{fig1}. (Fermion-loop
diagrams do not contribute to $\beta$ functions at 
the one loop level in both, the undeformed  and deformed models.
 We discuss them at the end of this section.)
For dimensional reasons we need each diagram to be at most  quadratic in $\partial^\mu\phi_0$.
Higher orders in this parameter will not contribute to the effective action (see Eq.~(\ref{fri5})).
As a  result we arrive at the following contributions to the
effective Lagrangian:
\beqn
(a)&=&
iG^{1,1}\partial^\mu\phi_0\partial_\mu\phi_0 I G^{-1}\,,
\label{mon10}\\[3mm]
(b)&=&
\frac{i}{2d}\{{G^{1,0}}^2\partial_\mu\phi\partial^\mu\phi_0+
{G^{0,1}}^2\partial_\mu \phi^\dagger_0\partial^\mu \phi^\dagger_0\}I G^{-2}\nonumber\\[3mm]
&&
-\frac{i}{2}G^{1,0}G^{0,1}\left(\partial_\mu\phi_0\partial^\mu\phi^\dagger_0\right)I\, G^{-2}
\label{mon12}\,,
\eeqn
where we define the integral ($d=2-\epsilon$)
 \beq
 I=\int\frac{d^d p}{(2\pi)^d}\,\frac{1}{p^2-m^2} = -\frac i{4\pi} \Gamma\left(\frac{\epsilon}2\right)\, \left(\frac{\sqrt{4\pi}}{m}\right)^\epsilon=-\frac i{2\pi\epsilon}+O(\epsilon^0) \,.
 \label{mon13}
 \eeq
We will have many encounters with this universal integral later.

 Using  $\phi_0$ from Eq.~(\ref{fri4})
and assembling Eq.~(\ref{mon10}) -- (\ref{mon12}) we arrive at the following effective one-loop Lagrangian:
\beqn
\Delta{\mathcal L}_1
&=&
ik^2\,I\,\left[G^{1,1}Gf^\dagger f-\frac{1}{4}\left({G^{1,0}}\right)^2f^2-
\frac{1}{4}\left({G^{0,1}}\right)^2 f^{\dagger 2}-\frac{1}{2}G^{1,0}G^{0,1}f^\dagger f\right]G^{-2}
\nonumber\\[3mm]
&=&
-2ik^2f^\dagger fI\frac{1}{(1+f^\dagger f)^2}\,,
\label{mom14}
\eeqn
to be compared with Eq.~(\ref{fri5}). This one-loop result
obviously maintains the target-space symmetries. The integral $I$ can be readily calculated upon transition to the Euclidean space,
\beq
I=\frac{-i}{(4\pi)^{d/2}}\Gamma\left(1-\frac{d}2\right)\left(m^2\right)^{-1+\frac{d}2}\,,
\eeq
cf. Eq.~(\ref{mon13}). Finally, we arrive at
\beq
\Delta{\mathcal L}_1 =-\frac{2}{(1+\phi^\dagger\phi)^2}\partial^\mu\phi^\dagger\partial_\mu\phi\cdot\frac{1}{4\pi m^\epsilon}
\,\frac{2}{\epsilon}\,.
\eeq
Comparing with Eq.~(\ref{fri5}) it is easy to see that
\beq
\frac{1}{g^2}  
= \frac{1}{g^2_0}- i\,I =
\frac{1}{g^2_0}- \frac{1}{2\pi}\,\text{ln}\frac{M_{\rm uv}}{m}\,.
\eeq
Now, the general definition of the $\beta$ function 
\beq
\beta_g = \left(\partial/\partial {\rm ln}(m) \right) g^2(m)
\label{mon18}
\eeq
implies the following one-loop $\beta$ function:
\beq
\beta_g = -\frac{g^4}{2\pi}\,.
\label{betag1}
\eeq

We end this section by commenting on 
the fermion diagrams. In the  undeformed $\mathcal{N}=(2,2)$ model, the relevant diagram is 
the $T$ product of two 
 fermion U$(1)$ currents $\bar\psi\gamma^\mu\psi$ (Fig.~\ref{1loopf}), which is finite due to transversality. 
 Moreover, this
remains to be the case in the deformed $\mathcal{N}=(0,2)$ model too
--- all  relevant fermion diagrams have no
$\ln M_{\rm uv}/\mu$. A more detailed 
argumentation can be found in Appendix B.
Thus,  Eq.~(\ref{betag1}) presents the full contribution at one loop.

\section{$Z$-factors }
\label{section5}

The renormalization of the term $\mathcal{B}B A^\dagger (1+ A^\dagger A)^{-1}$  
is determined by one-particle irreducible corrections to the vertex  and the wave-function renormalization for $\psi_R$ and $\zeta_R$. Here we will deal with the corresponding $Z$ factors,
$Z_\zeta$ and $Z_{\psi_R}$ (for brevity the latter will be denoted as
$Z_{\psi}$).

A closer look at the Lagrangian given in Eqs.~(\ref{11}) and~(\ref{17}), tells us that 
this Lagrangian is invariant under a SU$(2)$ rotation of $\mathcal{B}$ and $\frac B {1+ A^\dagger A}$. 
If we define a SU$(2)$ superfield doublet
\beq
\Psi = \sma \frac {\sqrt{2}B}{g(1+ A^\dagger A)}\\ \mathcal{B}\smaa\,,
\eeq
the part 
that involves all right-handed fermions 
(i.e., all but the first terms in Eq.~(\ref{baresuperlagr})) can be rewritten as 
\beq
2\Psi^\dagger_a\Psi_a+\sqrt{2}\left[\frac\gamma g A^\dagger \varepsilon^{ab}\Psi_a\Psi_b
+{\rm H.c.}\right]\,,
\eeq
which is obviously SU$(2)$ invariant.\footnote{We would like 
to comment here that this symmetry does not commute with the target space symmetry, as the first component of the doublet is not invariant while the second component is.}
This symmetry has a discrete symmetry inside, which is not affected by fermion anomalies. 
So we conclude that the wave-function renormalization for $\psi_R\,\sqrt{\frac 2 {g^2\chi^2}}$ 
on the one hand, and for $\zeta_R$ on the other, must be the same,
$$Z_\zeta =Z_{\psi}\,.$$
In addition to the perturbative calculation presented below, 
this can also be indirectly deduced from Chen's analysis \cite{chen} of the fermion zero modes in the instanton background.

Now we are ready to carry out the actual calculation to verify the above expectation, and calculate the wave-function renormalizations for $\psi_R$ and $\zeta_R$. We start from the Lagrangian with the bare couplings in UV,
\beqn
 \mathcal{L}_{0}
&=& \int d^2\theta_R \frac 2{g_0^2}\left\{\frac{A\partial_R A^\dagger+ A^\dagger \partial_R A}{1+ A^\dagger A}+\frac{2 B^\dagger B}{(1+ A^\dagger A)^2}\right\} +2\mathcal{B}^\dagger\mathcal{B} \nonumber\\[3mm]
&&-\frac{4}{g_0^2} \left[ \frac{\gamma_0\, \mathcal{B} B A^\dagger}{1+ A^\dagger A}+ {\rm H.c.}
\right]\,,
\label{baresuperlagr}
\eeqn
and evolve it down,\footnote{At one loop there is no need distinguish between the Wilsonian 
action and the generator of the one-particle irreducible vertices, as they are the same. 
A detailed discussion can be found in \cite{SV86}.} where we have
\beqn
 \mathcal{L}_{\text {eff}}
&=& \int d^2\theta_R \frac 2{g^2}\left\{\frac{A\partial_R A^\dagger+ A^\dagger\partial_R A}{1+ A^\dagger A}+Z_\psi \frac{2 B^\dagger B}{(1+ A^\dagger A)^2}\right\} +2Z_\zeta \mathcal{B}^\dagger \mathcal{B} \nonumber\\[3mm]
&&-\frac{4}{g_0^2} Z_\gamma \left[ \frac{\gamma_0\, \mathcal{B} B A^\dagger}{1+ A^\dagger A}+ {\rm H.c.}
\right]\,.
\eeqn
Finally, we redefine the fields $\mathcal{B}$ and $B$ to absorb the $Z_\zeta$ and $Z_\psi$ factors.  
With this absorption done, we get
\beqn
\mathcal{L}_{\text {eff}}
&=& \int d^2\theta_R \frac 2{g^2}\left\{\frac{A\partial_R A^\dagger + A^\dagger \partial_R A}{1+ A^\dagger A}+\frac{2 B^\dagger B}{(1+ A^\dagger A)^2}\right\} +2\mathcal{B}^\dagger \mathcal{B} \nonumber\\[3mm]
&&-\frac{4}{g_0^2} \frac {Z_\gamma}{\sqrt{Z_\psi Z_\zeta}}\left[ \frac{\gamma_0\, \mathcal{B} B A^\dagger}{1+ A^\dagger A}+ {\rm H.c.}
\right]\,.
\label{Lgammaren}
\eeqn

\begin{figure}
\begin{center}
\includegraphics[width=2in]{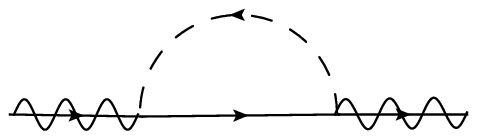}
\includegraphics[width=2in]{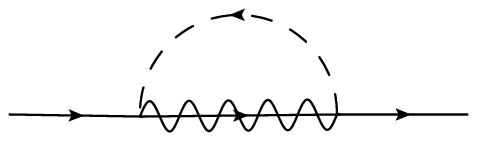}
\end{center}
\caption{\small One-loop wave-function renormalization of $\zeta_R$ and $\psi_R$. The solid line denotes the fermion field $\psi_R$. The solid line with a wavy line superimposed corresponds to the field $\zeta_R$. The fermion fields have their quantum parts and background parts marked by the same lines.}
\label{Zfactorppp}
\end{figure}

The one-particle irreducible diagrams for $Z_\zeta$ and $Z_\psi$ are shown in Fig.~\ref{Zfactorppp}. The calculation can be done by applying the linear background field method to $\psi_R$ and $\zeta_R$ separately, similarly to our description in Sec.~\ref{section4}. 
It is straightforward to check that
\beq
Z_\psi=Z_\zeta=1+i\gamma^2I\,,
\label{44}
\eeq
or, in other words,
\beq
\zeta_R\to \zeta_R\left(1+\frac i2\gamma^2I\right)\,,\qquad 
\psi_R\to \psi_R\left(1+\frac i2\gamma^2I\right)\,.
\eeq
We collect only divergent terms, where $I$ is given in Eq.~(\ref{mon13}). One can see that $\zeta_R$ and $\psi_R$ are corrected by $\gamma^2$ in the same way. In principle we still need to collect the $g^2$ correction to $\psi_R$. With the canonically normalized kinetic term
such a correction would be  absorbed in $g$. 
With our current normalization (see Eq.~(\ref{11})), it is simply absent.
Thus,  Eq.~(\ref{44})  is the full result for the wave-function renormalization. 

\section{One-loop $\beta$ function for $\gamma$}
\label{section6}

\begin{figure}
\begin{center}
\includegraphics[width=6in]{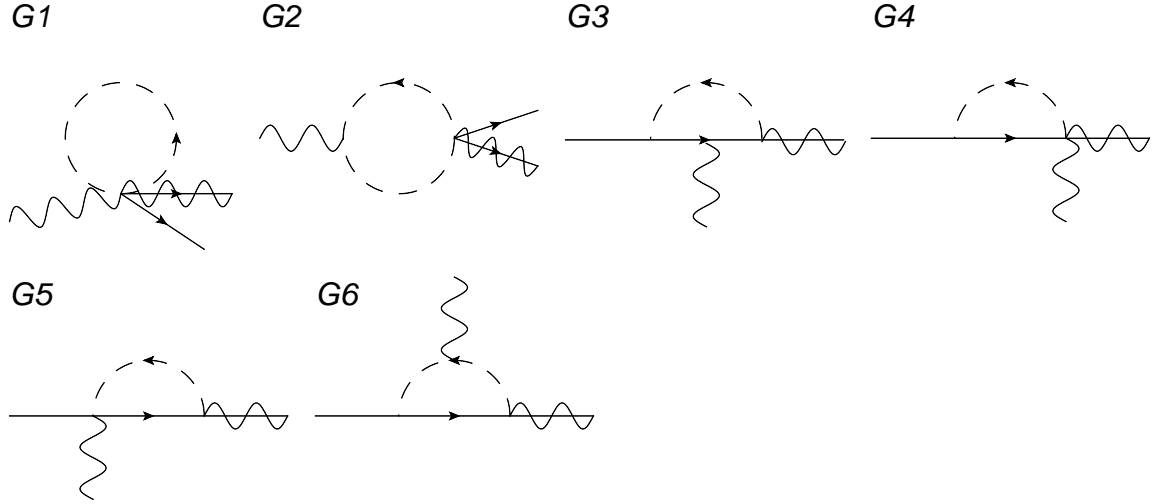}
\end{center}
\caption{\small One-loop 1PI diagrams contributing to the 
renormalization of  $\zeta_R R(i\partial_L\phi^\dagger)\psi_R$. Their overall sum
vanishes, see Table~\ref{gamma1table}.}
\label{gamma1}
\end{figure}

From Eq.~(\ref{Lgammaren}), we see that 
\beq
\frac \gamma{g^2}=\frac {Z_\gamma}{\sqrt{Z_\zeta Z_\psi}}
\,\,
\frac{\gamma_0}{g_0^2}\,,
\eeq
where $Z_\gamma$ collects all one-particle irreducible diagrams 
that correct the $\mathcal{B}B \bar A$ vertex.  In Sec.~\ref{section3} 
we used a general argument to show that the $O(g^2)$ correction to 
$Z_\gamma$ should vanish. Here we verify this statement, by explicit 
one-loop calculations. The relevant diagrams are shown in Fig.~\ref{gamma1}, 
and their contributions are listed in Table.~\ref{gamma1table}. It is seen that 
the total sum of all diagrams vanishes.

\begin{table}
\begin{center}
\begin{tabular}{| c || c |}
\hline
Diagram & Result\\[2mm]
\hline
G1 & $-\frac{2(2f^\dagger f-1)}{(1+f^\dagger f)^2}\gamma  \zeta_R\partial_L\phi^\dagger \psi_R I$ \\[2mm]
G2 & $\frac{8f^\dagger f}{d(1+f^\dagger f)^2}\gamma  \zeta_R\partial_L\phi^\dagger \psi_R I$ \\[2mm]
G3 & $-\frac{4f^\dagger f}{(1+f^\dagger f)^2}\gamma \zeta_R\partial_L\phi^\dagger \psi_R I$ \\[2mm]
G4 & $\frac{4f^\dagger f}{(1+f^\dagger f)^2}\gamma \zeta_R\partial_L\phi^\dagger \psi_R I$ \\[2mm]
G5 & $\frac{2(2f^\dagger f-1)}{(1+f^\dagger f)^2}\gamma \zeta_R\partial_L\phi^\dagger \psi_R I$ \\[2mm]
G6 & $-\frac{8f^\dagger f}{d(1+f^\dagger f)^2}\gamma \zeta_R\partial_L\phi^\dagger\psi_R I$ \\[2mm]
\hline
\end{tabular}
\end{center}
\caption{\small One-loop results for $ R \zeta_Ri\partial_L\phi^\dagger\psi_R $ ($1/\epsilon$  terms
coming from the integral $ I$).}
\label{gamma1table}
\end{table}

\begin{figure}
\begin{center}
\includegraphics[width=4in]{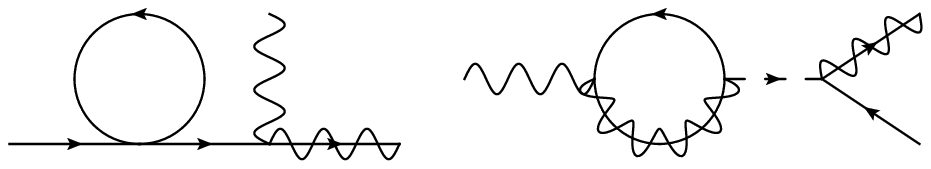}
\end{center}
\caption{\small Examples of one-loop diagrams  $O(\gamma^2)$, which do not 
contribute to the low-energy effective action since they are one-particle reducible. }
\label{trivialgamma}
\end{figure}

As for $O(\gamma^2)$ correction to the $\mathcal{B}B \bar A$ vertex, at one
loop the would-be  relevant diagrams either contain three vertices linear in $\gamma$, or involve one four-fermion interaction which contributes $O(\gamma^2)$. In either case, these 
diagram are one-particle reducible, as shown in Fig.~\ref{trivialgamma}. 
So, now it is verified that $Z_\gamma=1$, and, hence,
\beq
\gamma=\left(1-i\gamma^2I+ig^2I\right)\gamma_0\,,
\eeq  
implying in turn
\beq
\beta_\gamma=\frac {\gamma} {2\pi}(\gamma^2-g^2)\,.
\eeq

From symmetry arguments one should be convinced that the 
correction to the four-fermion interaction (the second line in Eq.~(\ref{N02model}))
must be totally determined by the wave-function 
renormalization of $\zeta_R$ and $\psi_R$ and by renormalization of $g^2$ and $\gamma$. 
As a consistency check we present 
the calculation of these terms in Appendix C, to show that it is precisely the case.

\section{Discussion on the running of the couplings}
\label{section7}

 Now that we know  how the  couplings $g^2$ and $\gamma$
 run in the leading order
 we can discuss the evolution of the two-coupling  theory at hand from the UV to IR or vice versa.
 
 The running of $g^2$ does not change (see Eq.~(\ref{betafsp})), it is still in the AF
 regime, much in the same way as in the undeformed ${\mathcal N}=(2,2)$
 model. 
Given the definition of the deformation constant in Eq.~(\ref{N02model}),
we see that of interest is the evolution law of the ratio
\beq
\rho(\mu)=\left(\frac{\gamma(\mu)}{g(\mu)}\right)^2\,.
\label{45}
\eeq
This is the coupling constant in front of the four-fermion term.
 As we will see shortly, to avoid the Landau pole in the UV 
 we must choose 
 $$\rho_0=\rho(M_{\rm uv})\leq\rho_*\equiv \frac{1}{2}\,.$$
Indeed, assembling together our results for the $\beta_{g,\gamma}$ functions we find
 \beq
\beta_\rho =\frac {g^2}{2\pi}\rho(2\rho-1)\,.
\label{47}
\eeq
 If $\rho<\rho_*$, the $\beta$ function is negative
 implying the AF regime. If, on the other hand, $\rho>\rho_*$, the $\beta$ function is positive,
 implying the 
 existence of the 
 Landau pole at a large value of the
 normalization point. The boundary value $\rho_*=1/2$ is a fixed point. If at an
 intermediate  normalization point\,\footnote{$\mu$ cannot be too small.
 It must be large enough to guarantee that $g^2(\mu) < 1$. Otherwise we are
 outside the domain of perturbation theory.} $\mu$ we choose
 $\rho <\rho_*$, it will run according to the AF law in UV and will tend to $1/2$ in IR.
 
\begin{figure}
\begin{center}
\includegraphics[width=4in]{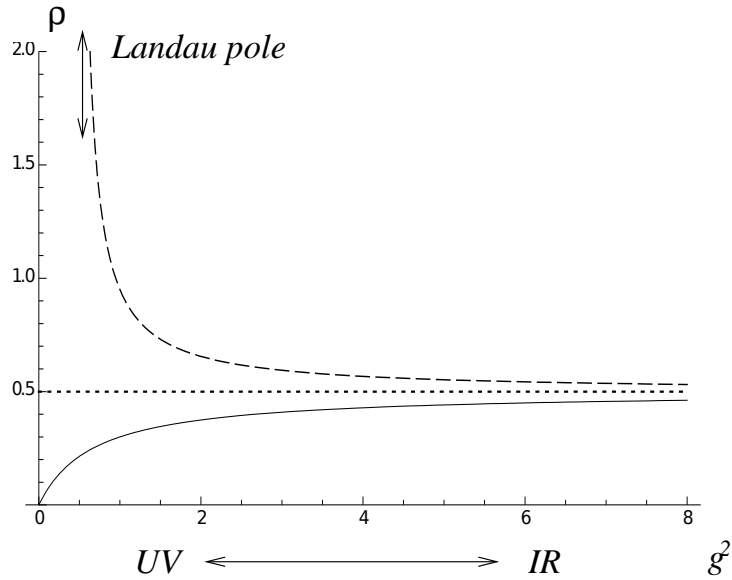}
\end{center}
\caption{\small $\rho$ versus $g^2$. The dashed, dotted and solid lines correspond to the cases $c<0$, $c=0$, $c>0$ respectively.}
\label{Rho-g2}
\end{figure}

 Actually,
 it is simple to find 
 an analytic solution for $\rho$ as a function of $g^2$, by rewriting Eq.~(\ref{47}) as
 \beq
 \frac{d\rho}{\rho(2\rho-1)}=-d\left(\text{ln}(g^2)\right),
 \label{51}
 \eeq
where on the right-hand side we used Eq.~(\ref{betag1}).
Eq.~(\ref{51}) implies
 \beq
 \rho(g^2)=\frac{1}{2+\frac {c}{g^2}}\,,
 \label{rho_running}
 \eeq
 where the constant $c$ is fixed by the boundary conditions. 
 In Fig.~\ref{Rho-g2} we observe  a universal  IR behavior of $\rho$ approaching $1/2$ from both sides.
Of course, so far our derivation was purely perturbative, and our result 
was obtained at one loop, which formally precludes us from 
penetrating too far in the infrared, where the coupling $g^2$ explodes.
However, later we will argue that IR the fixed point at $\rho=1/2$
survives beyond this approximation.

Figure~\ref{Rho-g2} also exhibits the pattern of the UV behavior.
If $c$ is positive, $\rho$ is asymptotically free. For negative $c$
one hits the Landau pole at a large (but finite) value of the normalization point.
Below we will not consider this regime because of its inconsistency.
If we take $c=0$  the value for $\rho$ freezes at $\rho_*=1/2$
in the weak coupling domain, and, formally, remains at $1/2$ in the strong coupling domain too.
As we will see later, in the heterotic CP$(N-1)$
models with $N>2$, the boundary value $\rho=1/2$ is a special point where a certain 
chiral flavor symmetry is restored.

\section{Extension to CP$(N-1)$ with $N>2$}
\label{section8}

In this section, we 
will generalize our analysis to the  heterotic CP$(N-1)$  sigma model. We will 
show that the parameter $\rho$ does not scale with $N$, and 
for any $N$ the function $\beta_\rho$ is the the same as in Eq.~(\ref{47})
at one loop. Then we will discuss
 a chiral fermion symmetry of the $\psi_L$'s sector at $\rho=1/2$,
 which will give us an argument  
that $\beta_\rho$ is proportional to $2\rho -1$ in all loops.

The heterotic CP$(N-1)$ Lagrangian in
the geometric formulation can
be borrowed from \cite{SY1}, 
\beqn
\mathcal{L}_{\text{CP}(N-1)}&&=G_{i\bar j}\left[\partial^\mu\phi^i\partial_\mu\phi^{\dagger \bar j}+
i\bar\psi^{\bar j}\slashed{D}\psi^i\right]\nonumber\\[3mm]
&&+i\zeta_R^\dagger\partial_L\zeta_R+\left[ \gamma \zeta_R G_{i\bar j}\left(i\partial_L\phi^{\dagger\bar j}\right)\psi_R^i+\text{H.c.}\right]+\gamma^2\left(\zeta_R^\dagger \zeta_R \right)\left( G_{i\bar j}\psi_L^{\dagger\bar j}\psi_L^i\right)\nonumber\\[3mm]
&&-\frac {g^2}2\left( G_{i\bar j}\psi_R^{\dagger\bar j}\psi_R^i\right)\left( G_{k\bar m}\psi_L^{\dagger\bar m}\psi_L^k\right)\nonumber\\[3mm]
&&+\frac {g^2}2\left(1-2\frac{\gamma^2}{g^2}\right)\left( G_{i\bar j}\psi_R^{\dagger\bar j}\psi_L^i\right)\left( G_{k\bar m}\psi_L^{\dagger\bar m}\psi_R^k\right)\,.
\label{CPNL}
\eeqn
We again apply the phase rotation of $\gamma$ to make it real. Moreover, $G_{i\bar j}$
in Eq.~(\ref{CPNL}) is the standard K\"aler metric in the Fubini--Study form. It is seen that
$\rho=1/2$ is a special value nullifying the last line in Eq.~(\ref{CPNL}).
The scaling of the coupling constants with $N$ is as follows \cite{BSY3}:
\beq
g^2\sim N^{-1}\,,\quad \gamma^2\sim N^{-1}\,,\quad \rho \sim N^0\,.
\label{54}
\eeq
Now we will establish that the $\beta$ functions $\beta_{g,\gamma}$
are compatible with the above scaling.

Our 
 background field 
 strategy 
 can still be applied here much in the same way as in CP(1). 
As well known,  in the undeformed model one has (e.g. \cite{Morozov:1984ad})
\beq
\beta_g=-\frac {Ng^4}{4\pi}\,.
\label{53}
\eeq 
As was demonstrated in Sec.~\ref{section4} (see also Appendix B),
 $\beta_g$ remains intact at one loop 
in the deformed model. Parallelizing our previous analysis it is not difficult
to get that 
\beq
Z_\zeta=1+i(N-1)\gamma^2 I\,,\qquad Z_\psi=1+i\gamma^2 I\,,
\eeq 
implying
\beq
\beta_\gamma=\frac{N\gamma} {4\pi}\left(\gamma^2-g^2\right)\,.
\label{55}
\eeq
Finally, combining Eqs.~(\ref{53}) and (\ref{55}) we conclude that in the heterotic CP$(N-1)$ model
\beq
\beta_\rho=\frac{Ng^2}{4\pi}\rho(2\rho-1)\,.
\label{58}
\eeq
Thus, the scaling laws Eq.~(\ref{54}) indeed go through Eqs.~(\ref{53}), (\ref{55}), and (\ref{58}).
Moreover, one can introduce the 't Hooft couplings
\beq
\lambda=Ng^2\,,\qquad \gamma'=\sqrt{N}\gamma\,, 
\eeq
in terms of which there are no explicit $N$ factors in the $\beta$ functions.
In particular,
\beq
\beta_\rho=\frac{\lambda}{4\pi}\rho(2\rho-1)\,,\qquad \rho = \frac{\lambda}{\left(
\gamma^\prime\right)^2}\,.
\eeq
All UV and IR regimes observed in CP(1) are maintained in the heterotic
$\mbox{CP$(N-1)$}$, in particular,
 the fixed point $\rho=1/2$. Note that the constant $c$ 
 expressing the boundary condition in the solution 
 Eq.~(\ref{rho_running}) must be rescaled in CP$(N-1)$, namely,
 \beq
 \rho(g^2)=\frac{1}{2+\frac {c}{Ng^2}}\,,
 \label{rho_runningp}
 \eeq

Now, after we dealt with
arbitrary values of $N$, it is time to turn to the issue of the
chiral fermion symmetry. A quick inspection of the
third and fourth lines in Eq.~(\ref{CPNL}) prompts us that their chiral structure is different.
The third line is invariant under {\em independent} SU$(N-1)$ rotations of $\psi_L^i$
and $\psi_R^j$, while the fourth line is not.\footnote{The SU$(N-1)$ rotations of $\psi_L^i$
 of which we speak here are rather peculiar since we deal with the 
nonflat target space. The fermions $\psi$ are defined on the tangent space, 
which is different for different points on the target manifold. Therefore, the SU$(N-1)$ rotations
cannot be global. For all terms other than kinetic, this is unimportant.
To properly define the chiral symmetry that leaves the kinetic term of $\psi_L$ 
invariant we have to impose an additional constraint. See Appendix D for more details.}
A class of rotations we keep in mind is the SU$(N-1)$ rotations of $\psi_L$'s, with
$\psi_R$'s, $\zeta_R$, and bosonic fields intact.
In the geometric formulation one introduces vielbeins $E^a_i$
and rotates 
\beq
E^a_i\,\psi^i_L \to U^a_b\, E^b_i\,\psi^i_L\,\qquad  E^a_i\,\psi^i_R\to E^a_i\,\psi^i_R\,,
\label{oxuel}
\eeq
where $U^a_b$ is a matrix from SU$(N-1)$.
A similar transformation law can be written in the gauge formulation \cite{SY1,BSY3} 
too.\footnote{In the notation of \cite{BSY3} 
the symmetry enhancement occurs at $|\tilde{\gamma}|^2=1$.} 
Note that in the CP(1) model the only possible chiral transformation of fermions 
is U(1), and it is anomalous. 
Hence, in fact there is no such symmetry. It starts from 
$N\geq 2$ in which case the chiral transformation is non-Abelian and nonanomalous.
Consideration of the
chiral fermion symmetries in the heterotic CP$(N-1)$ model was started by Tong~\cite{Tong07}.

At $\rho=1/2$ the fourth line in Eq.~(\ref{CPNL})
vanishes. Other terms in the Lagrangian are invariant under the
SU$(N-1)$ rotations of the left-handed fermions Eq.~(\ref{oxuel}). 
(For the kinetic term of $\psi_L$'s see the discussion in Appendix D.)

Thus, at $\rho =1/2$ the symmetry of the heterotic CP$(N-1)$ Lagrangian is enhanced.
It seems likely that this enhancement (and, hence, the fixed point at $\rho =1/2$ which goes with it)
will hold to all
orders in the coupling constant. 
Indeed, if one remembers about the origin of the
heterotic CP$(N-1)$ model as the world-sheet theory on the strings supported 
in $\mu A^2$ deformed ${\mathcal N}=2$ SQCD, one can try to 
relate the above symmetry enhancement at $\rho =1/2$
with that in  the bulk theory \cite{Tong07}. In the limit $\mu\to \infty$ 
the bulk theory becomes ${\mathcal N}=1$ SQCD  acquiring a chiral symmetry absent
at finite $\mu$. Remarkably, the $\mu\to \infty$ limit corresponds to
$\rho\to 1/2$ on the world sheet \cite{BSY1,BSY2}. Thus, we expect
the $\beta$ function $\beta_\rho$ to be proportional to $2\rho -1$  to all orders.
We plan to explore this issue in more detail in \cite{CS2}.

Now we would like to argue that the solution Eq.~(\ref{rho_runningp}) is, in fact,  {\em valid to  all} orders
in perturbation theory in the (planar) limit of large $N$, and so is
Eq.~(\ref{58}) for $\beta_\rho$. Indeed, the heterotic CP$(N-1)$ model
was solved in the  large-$N$ (planar) limit \cite{largen}.
The heterotic deformation parameter
determining a number of physical quantities (e.g. the vacuum energy density)
is \cite{largen}
\beq
u = \frac{\rho}{ (1-2\rho)\,(Ng^2)}\,;
\label{62}
\eeq
it must be renormalization-group invariant. In addition, $u$ does not scale with $N$. Substituting the solution Eq.~(\ref{rho_runningp})
in Eq.~(\ref{62}) we indeed get
\beq
u = \frac{1}{c}\,,
\label{63}
\eeq
{\em quod erat demonstrandum}. 
The normalization-point independence and $N^0$ scaling law are explicit in Eq.~(\ref{63}).

\section{Conclusions}
\label{conclu}

In this paper we 
started the study  of  perturbation theory in the recently found $\mathcal{N}=(0,2)$ CP$(N-1)$ sigma models. We 
carried out explicit calculations of both relevant $\beta$ functions at one loop
and demonstrated that the theory is asymptotically free much in the same way as the unperturbed
$\mathcal{N}=(2,2)$ CP$(N-1)$ models provided the initial condition for $\gamma^2$
is chosen in a self-consistent way (i.e. $c$ is positive). The $\beta$ function for the ratio
$\rho =\gamma^2/g^2$ exhibits an IR fixed point at $\rho=1/2$.
Formally this fixed point lies outside the validity of the one-loop
approximation. We argued, however, basing on additional considerations,
that the fixed point at $\rho =1/2$ may survive to all orders.
The reason is the enhancement of symmetry (restoration of a chiral fermion flavor
symmetry) at $\rho =1/2$.
 Moreover, we argued that Eq.~(\ref{rho_runningp}) for $\beta_\rho$
 formally obtained at one loop, is in fact {\em exact to all orders} in the
 large-$N$ (planar) approximation. Thus, in this approximation the fixed point
 at $\rho=1/2$ is firmly established.

In addition to  the 
above quantitive results, we also got insights on field-theoretical aspects of the heterotic model. 
Using the $\mathcal{N}=(0,2)$ superfield language, we saw that in CP(1) both fields 
$\psi_R$ and $\zeta_R$ get the same renormalization at one loop level. This is due to  an 
unexpectated  and unusual SU$(2)$ symmetry between $\psi_R$ and $\zeta_R$.  The
novelty of this symmetry is quite obvious
because it mixes  chiral  and anti-chiral superfields, and does not commute with the
target space symmetry. 

\section*{Acknowledgments}

XC thanks N. Seiberg for very helpful discussions. 
MS is grateful to A. Yung for endless illuminating discussions.
Useful remarks due to S. Bolognesi and D. Tong are gratefully acknowledged.

This work was completed during an extended visit of MS to the Jefferson Physical Laboratory, Harvard University. I would like  to thank  my colleagues for hospitality.
The work of MS was supported in part by DOE grant DE-FG02-94ER408.

\begin{appendix}

\addcontentsline{toc}{section}{Appendices}
\section*{Appendix A}
\renewcommand{\theequation}{A.\arabic{equation}}
\setcounter{equation}{0}

In this appendix we  describe the $\mathcal{N}=(0,2)$ superspace in  $D=1+1$ dimensions, present our notation, 
and   derive Eq.~(\ref{n02transformation}). 

The space-time coordinate $x^\mu=\{t,z\}$ can be promoted to superspace by adding a complex Grassmann variable $\theta_R$ and its complex conjugate $\theta_R^\dagger$. Where-ever our expressions are dependent on the representation of Clifford algebra, we use the following convention.
\beq
\gamma^0=\sma 0 & -i\\ i & 0\smaa\,,\qquad \gamma^1=\sma 0 & i\\ i & 0\smaa\,, \qquad\gamma^3=\gamma^0\gamma^1=\sma 1 & 0\\ 0 & -1 \smaa\,.
\eeq
Under this representation Dirac fermion is expressed as 
\beq
\psi=\sma\psi_R\\ \psi_L\smaa\,,
\eeq 
and we have 
\beq
\slashed{\partial}=\sma 0 & -i\partial_R\\ i\partial_L & 0\smaa\,,
\eeq
 where $\partial_R=\partial_t-\partial_z$ and $\partial_L=\partial_t+\partial_z$.
The corresponding supercharges are given by
\beq
Q_R=i\frac\partial{\partial \theta_R}-\theta_R^\dagger\partial_L\,,\qquad \bar{Q}_R=\frac\partial{\partial\theta_R^\dagger}+i\theta_R\partial_L\,.
\eeq

Applying them to the superspace, we have the transformation rules as follow.
\beqn
&&i(Q_R\epsilon_R+\bar{\epsilon}_R \bar{Q}_R)x^\mu=
i\epsilon_R^\dagger\theta_R-i\theta_R^\dagger\epsilon_R\,,\nonumber\\[3mm]
&&i(Q_R\epsilon_R+\bar{\epsilon}_R \bar{Q}_R)\theta_R=\epsilon_R\,,\qquad i(Q_R\epsilon_R+\bar{\epsilon}_R^\dagger \bar{Q}_R)\theta_R^\dagger=\epsilon_R^\dagger\,,
\label{trans1}
\eeqn
where $\mu\in\{0,1\}$, $\bar{\epsilon}_R=-i\epsilon_R^\dagger$.
So in fact if we define 
$$
x_L^\mu=x^\mu+i\theta_R^\dagger\theta_R\,\,\mbox{ and}\,\, x_R^\mu=x^\mu-i\theta_R^\dagger\theta_R\,,
$$
then we have 
\beq
i(Q_R\epsilon_R+\bar{\epsilon}_R \bar{Q}_R)x_L^\mu=2i\epsilon_R^\dagger\theta_R\,,\qquad i(Q_R\epsilon_R+\bar{\epsilon}_R \bar{Q}_R)x_L^\mu=-2i\theta_R^\dagger\epsilon_R\,.
\eeq 

Now we are ready to deduce the transformation law for the chiral superfield $A$ and $B$. 
\beqn
&&i(Q_R\epsilon_R+\bar{\epsilon}_R \bar{Q}_R)[\phi(x_L)+\sqrt{2}\theta_R\psi_L(x_L)]=\partial_L\phi 2i\epsilon_R^\dagger\theta_R+\sqrt{2}\epsilon_R\psi_L\,,\nonumber\\[3mm]
&&i(Q_R\epsilon_R+\bar{\epsilon}_R \bar{Q}_R)[\psi_R(x_L)+\sqrt{2}\theta_R F(x_L)]=\partial_L\psi_R 2i\epsilon_R^\dagger\theta_R+\sqrt{2}\epsilon_RF\,.
\eeqn
And this immediately leads us to Eq.~(\ref{n02transformation}).

Finally, we comment that for the chiral superfield $A$, we have that $\partial_R A$ is also a chiral superfield. This is because of the fact that $[Q_R, \partial_R]=[\bar{Q}_R, \partial_R]=0$.

\section*{Appendix B}
\renewcommand{\theequation}{B.\arabic{equation}}
\setcounter{equation}{0}

\begin{figure}
\begin{center}
\includegraphics[width=1.7in]{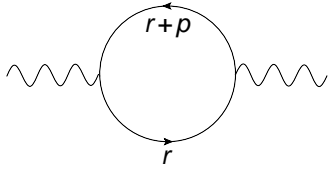}
\end{center}
\caption{\small One-loop contribution by fermionic loop. The solid lines denote fermionic propagators, and the wave lines are bosonic background fields. }
\label{1loopf}
\end{figure}

In the limit of small $m$ the one-loop fermion contribution is determined by the diagram depicted in Fig.~\ref{1loopf}.
Let us start from $m=0$. Then the fermion Green's function is
\beq
\left\langle T\, \psi (x)\,\bar\psi (0)\right\rangle = -\frac{i}{2\pi}\frac{\slashed{x}}{x^2}\,.
\eeq
Correspondingly, 
\beqn
\left\langle T\, \bar\psi(x)\gamma^\mu \psi (x)\,\, \bar\psi (0)\gamma^\nu \psi(0) \right\rangle 
&=& \frac{1}{2\pi^2}\left(x^2 g^{\mu\nu} - 2x^\mu x^\nu
\right)\frac{1}{x^4}
\nonumber\\[4mm]
&\to&
\frac {i}{\pi}
\left(\frac{p^\mu p^\nu}{p^2} - g^{\mu\nu}
\right).
\label{tue22}
\eeqn
This expression is singular at $p\to 0$. However, if we keep a small IR regularizing mass, then
it must be multiplied by a function $f(p^2/m^2)$ which is proportional to
$p^2/m^2$ at small $p^2$. Thus the fermion loop in Fig.~\ref{1loopf} vanishes at $p\to 0$. 

\section*{Appendix C}
\renewcommand{\theequation}{C.\arabic{equation}}
\setcounter{equation}{0}

\begin{figure}
\begin{center}
\includegraphics[width=4in]{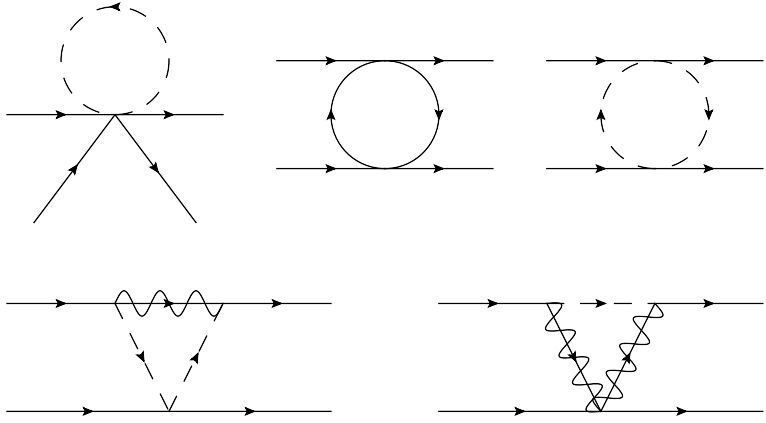}
\end{center}
\caption{\small One-loop corrections to $(\bar\psi\psi)^2$ term.}
\label{0614psi41loop}
\end{figure}

In order to simplify the calculation, we only collect the covariant contribution, and take the target space symmetry of the theory for granted. First we calculate one-loop correction to $(\bar{\psi}\psi)^2$. The relevant diagrams are shown in Fig.~\ref{0614psi41loop}.
Finally we have 
\beq
\Delta\mathcal{L}_{(\bar\psi\psi)^2}=
i\frac 2{(1+\phi^\dagger \phi)^4}(-1+\frac{\gamma^2}{g^2}+
\frac{\gamma^4}{g^4})I(\bar\psi \psi)^2\,.
\label{0614psi^4Lag}
\eeq

In order to see whether there are new structures, we recall that previously we have 
\beq
\mathcal{L}_{0,(\bar\psi\psi)^2}=\frac 2{g_0^2(1+\phi^\dagger\phi)^4}(\bar\psi_0\psi_0)^2-\frac {2\gamma_0^2}{g_0^4(1+\phi^\dagger\phi)^4}(\bar\psi_0\psi_0)^2\,.
\eeq 
So if the structure remains the same after one-loop renormalization, we should have 
\beqn
\mathcal{L}_{\text{eff},(\bar\psi\psi)^2}&=&\frac 2{g^2(1+\phi^\dagger\phi)^4}(\bar\psi\psi)^2-\frac {2\gamma^2}{g^4(1+\phi^\dagger\phi)^4}(\bar\psi\psi)^2\nonumber\\[3mm]
&=&Z_{g^2}^{-1}Z_\psi\frac 2{g_0^2(1+\phi^\dagger\phi)^4}(\bar\psi_0\psi_0)^2-Z_{\frac {\gamma}{g^2}}^2Z_\psi\frac {2\gamma_0^2}{g_0^4(1+\phi^\dagger\phi)^4}(\bar\psi_0\psi_0)^2\,,
\label{Leff_4f}
\eeqn
where 
\beqn
&&Z_{g^2}=\frac{g^2}{g_0^2}=1+iIg^2\,,\qquad Z_{\frac {\gamma}{g^2}}=\frac{\gamma g_0^2}{\gamma_0 g^2}=1-iI\gamma^2\,,\nonumber\\[3mm]
&&Z_\psi=1+iI\gamma^2\,.
\label{Zs_4f}
\eeqn 
The last one is because $\bar\psi\psi$ is linear in $\psi_R$(or $\psi_R^\dagger$). So if we plug in Eq.~(\ref{Leff_4f}) the known result Eq.~(\ref{Zs_4f}), we should expect that 
\beq
\Delta\mathcal{L}_{(\bar\psi\psi)^2}=i\frac 2{(1+\phi^\dagger\phi)^4}(-1+\frac{\gamma^2}{g^2}+\frac{\gamma^4}{g^4})I(\bar\psi\psi)^2\,.
\eeq
And it is precisely the case.

\begin{figure}
\begin{center}
\includegraphics[width=4in]{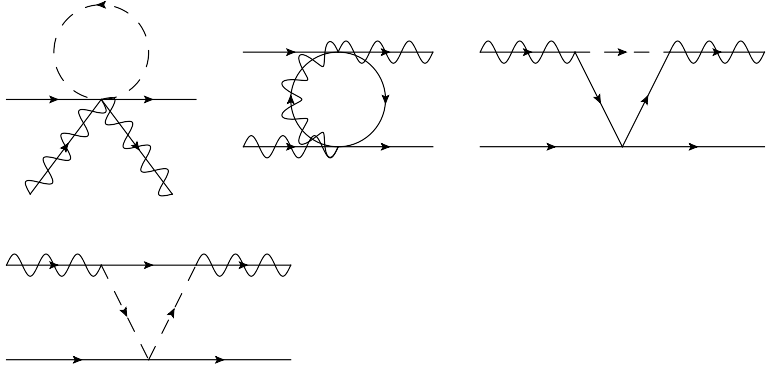}
\end{center}
\caption{\small One-loop corrections to $\zeta_R^\dagger\zeta_R\psi_L^\dagger\psi_L$ term.}
\label{0614zeta41loop}
\end{figure}

We can also calculate one-loop correction to $\zeta_R^\dagger\zeta_R\psi_L^\dagger\psi_L$, and the relevant diagrams are given in Fig.~\ref{0614zeta41loop}.
Finally we have 
\beq
\Delta\mathcal{L}_{\zeta_R^\dagger\zeta_R\psi_L^\dagger\psi_L}=i\frac 2{(1+\phi^\dagger\phi)^2}\frac{\gamma^2}{g^2}(g^2-\gamma^2)I\zeta_R^\dagger\zeta_R\psi_L^\dagger\psi_L\,.
\eeq
Following a similar analysis one can show that this is also consistent with our expectation.

\section*{Appendix D}
\renewcommand{\theequation}{D.\arabic{equation}}
\setcounter{equation}{0}

In this appendix we  introduce vielbeins $E_i^a$,  used in Sec.~\ref{section8} to describe the flavor symmetry of the left-handed fermions $\psi_L$.   The $\psi$ 
fermions   live in the tangent space of the target manifold. Vielbeins make it clear that locally we could find a coordinate frame  that is as good as the one for flat spaces. However, such choice of coordinates varies from point to point,  so one should expect that the fermions have nontrivial connections. 

Let us look at the fermionic part of the Lagrangian, which is given by Eq.~(\ref{CPNL}).  
It is convenient to write the K\"ahler metric as
$G_{\bar j i}$. Then all   vectors that carry the barred indices must be understood as lines, and all that carry the unbarred indices as columns. We consider the following
representation  of the metric: 
\beq
\left(E^\dagger\right)^a_{\bar j}E_{a,i}=G_{\bar j i}\,.
\eeq
Rising or lowering of the $a$ index  is done by the identity matrix, so we can be loose about its position. The above equation does not uniquely determine  the matrix $E$. Rather, we start 
with $(N-1)\times (N-1)$ complex matrix and impose $(N-1)^2$ real conditions. The remaining freedom  (the ambiguity can be represented by a constant U$(N-1)$ matrix) is non-physical and one can fix the ambiguity by imposing further compatible conditions. After   that, we can define our ``flat" fermions as 
\beq
\psi^a=E_i^a\psi^i\,.
\label{vielbeinferm}
\eeq

On the other hand, in any case the kinetic term for fermions needs some adjustments. 
Previously we had $\slashed{D} \psi^i\equiv \slashed{\partial} \psi^i+\Gamma^i_{lk}\slashed{\partial}\phi^l \psi^k$, and now,
 in order to require the covariant derivatives to be the same 
 both before and after we apply Eq.~(\ref{vielbeinferm}), we must have 
\beq
\slashed{D}\psi^a=E_i^a \slashed{D}\psi^i\,.
\eeq
These conditions determine that
\beq
D_\mu\psi^a=\partial_\mu\psi^a+\partial_\mu(E^a_i)\psi^i-E_i^a\Gamma^i_{lk}\partial_\mu\phi^l\psi^k\,.
\eeq

Generally speaking, it is impossible to choose the set of the vielbeins $E_i^a$ to 
reduce $D_\mu\psi^a$ to $\partial_\mu\psi^a$. The 
 reason is  simple: the change of local coordinate frames 
from one point on the target space to another is not trivial. 
In a sence, the U$(N-1)$ symmetries in choosing $E_i^a$'s are similar to 
a gauge symmetry. 

The symmetry we demonstrate here, is seen by replacing $\psi^i$'s by $\psi^a$'s. Now the 
fermion part of the Lagrangian takes the form
\beqn
\mathcal{L}_{\text{CP}(N-1)}&&=i\psi_R^{\dagger a}D_L\psi_R^a+i\psi_L^{\dagger a}D_R\psi_L^a
\nonumber\\[3mm]
&&+i\zeta_R^\dagger\partial_L\zeta_R+\left[ \gamma \zeta_R \left(i\partial_L\phi^{\dagger\bar j} E^\dagger_{\bar j a} \right)\psi_R^a+\text{H.c.}\right]+\gamma^2\left(\zeta_R^\dagger \zeta_R \right)\left( \psi_L^{\dagger a}\psi_L^a\right)\nonumber\\[3mm]
&&-\frac {g^2}2\left( \psi_R^{\dagger a}\psi_R^a\right)\left( \psi_L^{\dagger b}\psi_L^b\right)\nonumber\\[3mm]
&&+\frac {g^2}2\left(1-2\frac{\gamma^2}{g^2}\right)\left( \psi_R^{\dagger a}\psi_L^a\right)\left( \psi_L^{\dagger b}\psi_R^b\right)\,,
\label{CPNLviel}
\eeqn
where $D_R$ and $D_L$ are defined from $D_\mu$ in the way  similar   to   the 
replacement of $\partial_\mu$  by $\partial_R$ and $\partial_L$. 

As was emphasized before, since $D_L$ and $D_R$ do not reduce to $\partial_L$ and $\partial_R$, we cannot yet apply the flavor rotation to  $\psi_L^a$'s:
the corresponding kinetic term in the Lagrangian will not be invariant. But it will be invariant, if we further assume that $\phi$'s are only dependent on $t+z$. By doing so, $D_R \to \partial_R$, since the connection part is always linear in $\partial_R\phi^i$'s.
Now we can see that the last line
in Eq.~(\ref{CPNLviel}) is the only term that is noninvariant under the SU$(N-1)$ flavor rotation of $\psi_L^a$. Needless to say, the symmetry is restored when $\rho=1/2$.

\end{appendix}

\end{document}